\documentclass[twocolumn,showpacs,preprintnumbers,amsmath,amssymb]{revtex4}
\usepackage{graphicx}
\usepackage{dcolumn}
\usepackage{bm}

\def\mib#1{\mbox{\boldmath $#1$}}

\begin{document}

\title{
Boundary Dynamics of Sweeping Interface
}

\author{Hiizu Nakanishi}


\affiliation{
Department of Physics, Kyushu University 33, Fukuoka 812-8581, Japan
}

\begin{abstract}
A new type of boundary dynamics is proposed to describe the 
interface that sweeps space to collect distributed material.  Based upon
geometrical consideration on a simple physical process representing a
certain experiment, the dynamics is formulated as the small diffusion
limit of Mullins-Sekerka problem of crystal growth.  It is demonstrated
that a steadily extending finger solution exists for a finite range of
propagation speed, but numerical simulations suggest they are unstable
and the interface shows a complex time development.

\end{abstract}
\pacs{45.70.Qj, 81.10.Aj, 47.54.+r}

\maketitle

\section{Introduction}

The boundary dynamics has been providing interesting problems for
physics and mathematics.  A famous one is the problem of crystal growth
from a supercooled melt\cite{L80}; As the melt is solidified and a
crystal grows, a flat interface between the solid and the liquid phases
becomes unstable due to the coupling of the
solidification process with the diffusion of latent heat generated at
the interface (Mullins-Sekerka instability\cite{MS63-64}).  This
results in the fascinating variety of dendritic growths of
crystal under the interplay with anisotropic surface tension.  Another
example is the viscous fingering, which appears when the air is injected
into a viscous fluid (Saffman-Taylor instability \cite{BKLST86,ST58}).
The viscous fluid is displaced by the pressure gradient, and the
pressure field is governed by Laplace equation with proper boundary
conditions.  Since Laplace equation is the diffusion equation with the
infinite diffusion constant, the viscous fingering is the large
diffusion limit of the crystal growth.

A new example of boundary dynamics is presented by a simple experiment
of Yamazaki and Mizuguchi\cite{YM00};  The mixture of water and corn
starch powder is sandwiched between two glass plates. After a several
hours, a labyrinthine pattern of dried corn starch will be formed when
the water is evaporated from the gap of the glass plates.  The system is
two-dimensional, and the pattern is formed by the water-air interface
line as it sweeps the system to collect the granules along it by means
of the surface tension.  The grains are simply accumulated along the
interface line to give friction against the interface motion, and
eventually get stuck with the glass plates.

In the above experiment, the granules play an analogous role to that of
the latent heat in the crystal growth; They are
distributed in the wet (or melted) region, but show up when the
interface passes; The interface speed is controlled by the granule
density (or temperature) at the boundary.  The difference is that the
granules do not diffuse while the heat does.  In this sense, 
this {\em sweeping dynamics} is the small diffusion limit of
Mullins-Sekerka problem; That is the opposite limit to Saffman-Taylor
problem but has not been investigated yet in detail.

Analogous instabilities exist for these phenomena; A
protruded part of interface advances faster because the generated heat
diffuses faster  (Mullins-Sekerka), the pressure gradient is
larger (Saffman-Taylor), or the accumulated granules are diluted over
the elongated interface at the convex region (sweeping dynamics).

Similarities is seen in the phase field model proposed for this
sweeping phenomenon\cite{YMWM01,IHN05}.  The model consists of two
fields: the phase field and the coupling field; The coupling field
represents the granular density, instead of the temperature in the 
crystal growth\cite{L86,PF90,K93}.  It has been demonstrated the
model is capable to reproduce some feature of the patterns obtained in
the experiment\cite{IHN05}.

In this paper, we will construct the model of the boundary dynamics
for the sweeping interface based on geometrical and physical
considerations, and study its behavior.

\section{System configuration and co-ordinate}

\begin{figure}[b]
\begin{center}
\includegraphics[width=8cm,angle=0]{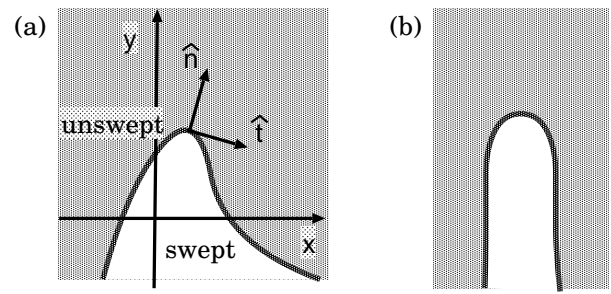}
\end{center}
\caption{
Schematic diagram of the interface between the swept and the unswept regions.
(a) Coordinate system with the normal and tangential vectors.
(b) Steadily extending finger.
}
\label{f-1}
\end{figure}

Let us start by defining the Cartesian coordinate in the two dimensional
system near the interface between the swept (dry) and the unswept (wet)
region as in Fig.\ref{f-1}(a).  In the unswept region, the granules are distributed
at the area density $\rho$, which we assume constant in this paper,
for simplicity.

Suppose that
the interface position at the time $t$ is represented parametrically by
\begin{equation}
  \mib r(t,s) = ( x(t,s), y(t,s) )
\label{vec_r}
\end{equation}
with the parameter $s$.
Let $\ell$ be the natural coordinate along the interface, then
the length element  $d\ell$ is
\begin{equation}
 d\ell = \sqrt{\left({\partial x /\partial s}\right)^2+
                  \left({\partial y/\partial s}\right)^2}
  \,\, ds \, .
\label{d-ell}
\end{equation}
We define the tangential and the normal unit vectors of the interface,
$\hat{\mib t}$ and $\hat{\mib n}$, by
\begin{equation}
\hat{\mib t} \equiv {\partial\mib r/\partial\ell}
 \, ,
\quad
\hat{\mib n} \equiv \hat{\mib z}\times\hat{\mib t} \, ,
\end{equation}
where $\hat{\mib z}$ denotes the unit vector along the $z$ axis
perpendicular to the system; The normal vector $\hat{\mib n}$ is
pointing into the unswept region.  The curvature $\kappa$ is defined as
\begin{equation}
\kappa \equiv -{\partial\hat{\mib t}\over\partial\ell}\cdot\hat{\mib n} \, ,
\end{equation}
which is positive when the interface is convex toward the unswept side.

In the drying process, 
the interface moves upwards.
Using an appropriate parameterization of $s$ in Eq.(\ref{vec_r}),
$\partial\mib r/\partial t$ can be made parallel to $\hat{\mib n}$, then
the normal speed of the interface motion $v_n$ is defined by
\begin{equation}
{\partial\mib r(t,s)\over\partial t} = v_n(t,s) \hat{\mib n}(t,s).
\end{equation}

\section{Dynamics of sweeping interface}

We now consider the sweeping dynamics where the granules are swept by the
interface;  They are accumulated along the interface and conveyed by it
in its normal direction.  If we ignore the
width of the region where the granules are accumulated, then the
accumulated granules are described by the line density $\sigma$
along the interface.

In the case where the granules are simply accumulated along the
interface and do not diffuse at all, the equation for $\sigma$ 
is determined geometrically and should be given by
\begin{equation}
{\partial\sigma(t,s)\over\partial t} =
 v_n(t,s) \Bigl[ \,\rho  -\kappa(t,s)\sigma(t,s) \Bigr] \,.
\label{eq-sigma}
\end{equation}
The first term of the right hand side simply represents the sweeping
accumulation.  The second term comes from the change of the interface
length as it advances; The length element along the interface increases
by the factor $(1+\kappa v_n \Delta t)$ during the short time period
$\Delta t$, thus the line density decreases by the factor of its
inverse.

The interface speed $v_n$, on the other hand, is given by the product of
the following two factors, i.e. the driving force to the interface, and
the mobility of the interface; (i) The driving force comes from the
pressure difference $\Delta P$ between the wet and the dry regions; As
the water evaporates, the volume of the wet region tends to shrink, then
the interface recedes due to the pressure difference.  When the
interface is curved, the driving force is given by the effective
pressure difference $(\Delta P-\gamma\kappa)$ with $\gamma$ being the
surface tension, thus is proportional to the factor $(1-a\kappa)$
with the capillary length $a\equiv\gamma/\Delta P$.
(ii) The mobility depends upon the granule density $\sigma$ and should
be a decreasing function of it because the granule friction with the
glass plates resists the interface motion.  The mobility becomes zero at
$\sigma_{\rm st}$ when the interface gets stuck with granules.

In the simplest case where no other scales of $\sigma$ are involved, we
can write down the equation for the interface speed as
\begin{equation}
v_n = v_0 f(\sigma/\sigma_{\rm st}) (1-a\kappa)
\label{eq-speed}
\end{equation}
with the characteristic speed $v_0$ and the dimensionless mobility
$f(x)$, which is a decreasing function with $f(0)=1$ and $f(1)=0$.

Eqs.(\ref{eq-sigma}) and (\ref{eq-speed}) define the boundary dynamics
of the sweeping interface, which shows morphological instability
analogous to the crystal growth dynamics; The part of the interface with
$\kappa>0$ tends to advance faster when $f'<0$.

The model contains four parameters, $\rho$, $v_0$, $\sigma_{\rm st}$,
and $a$, from which we can define the stuck-in distance $\ell_{\rm st}$
and the stuck-in time $t_{\rm st}$ by
\begin{equation}
\ell_{\rm st} \equiv {\sigma_{\rm st}/ \rho},\quad
t_{\rm st} \equiv {\ell_{\rm st}/ v_0},
\end{equation}
which are the distance and the time that the flat interface can
advances before it gets stuck.

In order to simplify the expressions, in the following, we employ
the dimensionless unit system where 
$\ell_{\rm st}=t_{\rm st}=\sigma_{\rm st}=1$,
then Eqs.(\ref{eq-sigma}) and (\ref{eq-speed})
are in the form of
\begin{eqnarray}
{\partial\sigma(t,s)\over\partial t} & = &
         v_n(t,s) \Bigl[  \, 1 -\kappa(t,s)\sigma(t,s)  \Bigr] \, ,
\label{eq-sigma-2}
\\
v_n(t,s) & = &  f\Bigl(\sigma(t,s)\Bigr) \Bigl( 1-R\kappa(t,s)\Bigr) \, ,
\label{eq-speed-2}
\end{eqnarray}
with the only one dimensionless parameter 
\begin{equation}
R\equiv {a/ \ell_{\rm st}} = {a\rho/ \sigma_{\rm st}} .
\end{equation}
This is the ratio of the capillary length to the stuck-in distance and
is a measure of the effect of the surface tension.

The functional form of the dimensionless mobility
$f(\sigma)$ in Eq.(\ref{eq-speed-2}) should reflect the physical mechanism
how the interface slows down and gets stuck due to the
accumulated granules. Here, we employ the simplest form
\begin{equation}
f(\sigma) = \left\{
\begin{array}{ll}
(1-\sigma) & \mbox{ for } 0\le \sigma \le 1 \\
0 &  \mbox{ for } \sigma > 1 
\end{array}
\right.   .
\label{f-function}
\end{equation}

\section{Simple solutions}

Now, we study the interface dynamics based on
Eqs.(\ref{eq-sigma-2}),(\ref{eq-speed-2}), and (\ref{f-function}).

\subsection{flat interface}
For the flat interface,  $\kappa=0$, thus the solution is easily obtained as
\begin{equation}
v_{n}(t) = e^{-t} \equiv v_{\rm f}(t),
\quad
\sigma(t) = 1-e^{-t} \equiv \sigma_{\rm f}(t).
\label{flat}
\end{equation}
The flat interface can advance only by $\ell_{\rm st}$(=1) before it gets
stuck because it simply accumulates material.

\subsection{steadily extending finger}

A possible mode of steady advance is the extending one-dimensional
finger of the width of $2\ell_{\rm st}$ by shoving the granules
asides(Fig.\ref{f-1}(b)).

This type of steadily extending finger solution can be obtained as follows.
Suppose the finger is extending in the $y$ direction with the speed
$V$, then Eqs.(\ref{eq-sigma-2}) and (\ref{eq-speed-2}) become
\begin{eqnarray}
-V\sigma' y'  & = & ( 1 -\kappa\sigma )  V x' ,
\label{steady-1}
\\
V x' & = &  (1-\sigma) ( 1- R\kappa ) ,
\label{steady-2}
\end{eqnarray}
where the primes denote the derivative by the natural coordinate $\ell$.
Then, $x'$ and $y'$ are related as
\begin{equation}
x'^2 + y'^2 = 1 ,
\label{steady-3}
\end{equation}
thus, Eqs. (\ref{steady-1}) -- (\ref{steady-3}) can be
solved for $x$, $y$, and $\sigma$ as functions of $\ell$ under the
physical boundary conditions
\begin{eqnarray*}
(x,y) = (0,0) & \mbox{ at } & \ell=0
\\
(x,y) = (\mp 1,-\infty),\quad \sigma=1 & \mbox{ at } & \ell=\pm\infty ,
\end{eqnarray*}
for a given set of $R$ and $V$.

For a given $R$, a steady solution is possible for a finite range
of $V$;  The condition at the tip that $x'=1$ and $y'=0$ at  $\ell=0$
in Eqs.(\ref{steady-1}) and (\ref{steady-2})
leads to the allowed range of $V$,
\begin{equation}
V\le (1-\sqrt R)^2 .
\end{equation}

\begin{figure}[b]
\begin{center}
\includegraphics[width=8cm,angle=0]{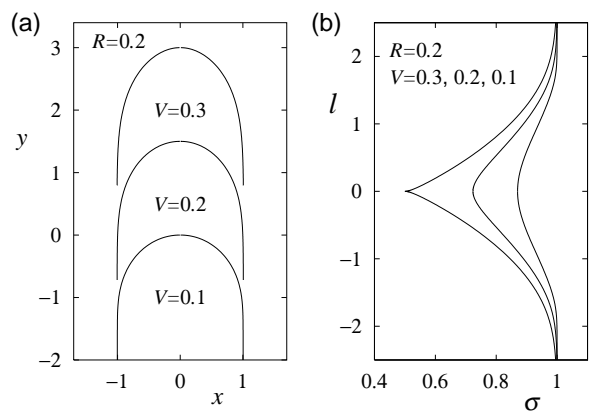}
\end{center}
\caption{ Steadily propagating finger solutions without diffusion for
$R=0.2$ with $V=0.3$, 0.2, and 0.1.  (a) Shapes of the finger. The plots
for different speeds are shifted to avoid overlapping.  Only the part
near the tip is shown for each finger.  (b) The line density of granules
v.s. $\ell$.  The finger tip is located at $\ell=0$.  } \label{f-2}
\end{figure}

Eqs.(\ref{steady-1})--(\ref{steady-3}) can be solved numerically.
Fig.\ref{f-2} shows some of the steady solutions for the finger shape
and the line density $\sigma$ for $R=0.2$ with $V=0.1$, 0.2, and 0.3.
One can see that the finger tip is sharper for the larger extending
speed $V$.

The line density $\sigma$ shows rather intriguing behavior as a function
of $\ell$; The tip density is smaller for the larger $V$, but for $V$
larger than a certain value for a given $R$, $\sigma$ becomes singular
at the tip and eventually develops a cusp.  Actually, by expanding
$\sigma$ around the tip, it can be shown that $\sigma$ becomes
non-analytic
\begin{equation}
\sigma(\ell)-\sigma(0) \propto \ell^\varphi
\end{equation}
with
\begin{equation}
\varphi = {2 \sqrt{(1-V+R)^2-4R}\over (1-V-R)-\sqrt{(1-V+R)^2-4R}}
\end{equation}
when
\begin{equation}
1+{1\over 3}\left[ 5R-4\sqrt{R^2+3R} \right] < V < (1-\sqrt R)^2  .
\label{sing-range}
\end{equation}
The singularity at the tip in the line density is a peculiar result of
the model without diffusion.  

\section{Ultra-violet catastrophe in the diffusionless model}

We have derived the steady finger solutions, which are smooth and
analytic except at the finger tip, but in a general time development,
the diffusionless feature of the model seems to cause the ultra-violet
catastrophe, or the short wave length instability, even though we have
taken account of the surface tension effect by introducing the capillary
length.

Actually, numerical simulations of the equations eventually yield a
zigzag structure of the solution in the shortest discretization length;
This cannot be accepted as a solution of the differential equations.

Mathematically, the problem is that the surface tension effect in
eq.(\ref{eq-speed-2}) is of the same form as $\kappa$ with the term that
causes the instability in eq.(\ref{eq-sigma-2}), thus the surface
tension effect never dominates to suppress the instability even in the
short wave length limit, unlike in the case of
Mullines-Sekerka/Saffman-Taylor problems.

This may be seen in the linear stability analysis of the flat interface
solution (\ref{flat})\cite{K05}.  Since the flat interface solution is
not steady, the linearized equations for the small deviation from it are
not of constant coefficient, thus the analysis is not simple, but within
the approximation where the time variation is neglected during the time
scale of the perturbation (quasi-steady approximation), the growth rate
$\Omega(q)$ of the unstable perturbation with the wave number $q$ in the
unstable branch can be obtained as
\begin{equation}
\Omega(q) = \frac{1}{2}
   \left[
     -(Rv_{\rm f}q^2-f_{\rm f}')+
\sqrt{(Rv_{\rm f}q^2-f_{\rm f}')^2-4f_{\rm f}'\sigma_{\rm f} v_{\rm f}q^2}
   \right] ,
\end{equation}
where $f_{\rm f}'=f'(\sigma_{\rm f}(t))$; $v_{\rm f}$ and $\sigma_{\rm f}$ are given by
eq.(\ref{flat}).
This is positive for any $q>0$ when $f_{\rm f}'<0$ and
\begin{equation}
\Omega(q) \to -\frac{f_{\rm f}'\sigma_{\rm f}}{R}
\quad \mbox{for }q\to\infty ,
\end{equation}
suggesting the ultra-violet catastrophe, thus the model is not well
defined in the continuum limit.

\section{Diffusion driven by interface motion}

In order to avoid this instability, we introduce the
diffusion term along the interface in Eq.(\ref{eq-sigma-2}) as
\begin{equation}
{\partial\sigma\over\partial t}  = 
         v_n \Bigl[ \, 1 -\kappa\sigma  \Bigr]
       + {\partial\over\partial\ell}\left(\ell_D v_n
          {\partial\sigma\over\partial\ell}\right)
\label{eq-sigma-3}
\end{equation}
with a new small length scale $\ell_D$.  The adopted form represents the
diffusion flux along the interface; The flux is proportional to the
interface speed $v_n$.  Such a diffusion flux is natural in the case
where the diffusion is driven by the interface motion; The grains
are driven randomly along the interface by the distance $\ell_D$ during
the time $\ell_D/v_n$, i.e.  the time during which the interface moves
by $\ell_D$ in the normal direction.  Note that we do not assume the
diffusion perpendicular to the interface.


The linear stability analysis within the same approximation as above
gives the perturbation growth rate
\begin{eqnarray}
\lefteqn{
\Omega(q) =
\frac{1}{2}\left[
-\Bigl((R+\ell_D)v_{\rm f} q^2 -f_{\rm f}'\Bigr) +
\right.}
\\
& & \left.
   \sqrt{\Bigl((R+\ell_D)v_{\rm f} q^2 -f_{\rm f}'\Bigr)^2+
       4\Bigl( (-f_{\rm f}')\sigma_{\rm f}-\ell_D Rv_{\rm f} q^2 \Bigr) v_{\rm f} q^2 }
\right],
\nonumber\end{eqnarray}
which gives
\begin{equation}
\Omega(q) < 0 \qquad \mbox{for }
 q>q_s\equiv\sqrt{\frac{(-f_{\rm f}')\sigma_{\rm f}}{\ell_D R v_{\rm f}}} ,
\end{equation}
therefore the catastrophe is suppressed.

\begin{figure}[b]
\begin{center}
\includegraphics[width=8cm,angle=0]{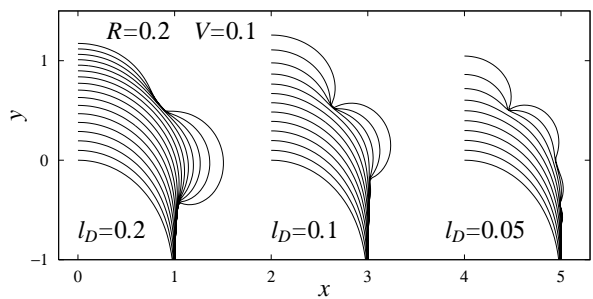}
\end{center}
\caption{ The time developments of the interface for $R=0.2$ with
$\ell_D=0.2$, 0.1, and 0.05.  Only the right halves of the fingers are
shown.  The time sequences are shown with the time interval $\Delta
t=1$.  The steady solution for $V=0.1$ with $\ell_D=0$ are used as the
initial configurations for all the cases.  } \label{f-3}
\end{figure}
%
%

\section{Numerical simulation of finger solutions}

I have performed simulations on the model with this diffusion
in order to see if the finger solutions are stable.

Let us start by examining the effect of the length scale $\ell_D$.
Fig.\ref{f-3} shows the results of simulations of Eqs.(\ref{eq-speed-2})
and (\ref{eq-sigma-3}) for $R=0.2$ with $\ell_D=0.2$, 0.1, and 0.05.
The steady solution of $V=0.1$ without the diffusion ($\ell_D=0$) is
used as the initial state.  Only the right halves of the interfaces are
shown and the time development is represented by the plots with the time
interval $\Delta t=1$.  I have confirmed the ultra-violet catastrophe is
suppressed by the diffusion term; Except for the cusps at the edges of
sticking regions, numerical solutions converges to a smooth solution as
the smaller time and space discretization is used for integral in
contrast with the diffusionless model, where the zigzag structure in the
smallest discretization scale develops eventually all over the interface
due to the ultra-violet catastrophe.  In all cases of Fig.\ref{f-3}, the
fingers extend at a speed of 0.1 initially with keeping the initial
shape as the steady solution does, but eventually the finger becomes
unstable and develops wavy structures.  The length scale of the emerging
structure is shorter for the smaller $\ell_D$, but they are much larger
than $\ell_D$.  We do not see any tendency that the steady solution
becomes stable for small $\ell_D$ even though the initial configuration
of the steady solution for the speed $V=0.1$ with $R=0.2$ is outside the
range Eq.(\ref{sing-range}), thus does not have a singularity in
$\sigma$ at the tip.

\begin{figure}[bt]
\begin{center}
\includegraphics[width=8cm,angle=0]{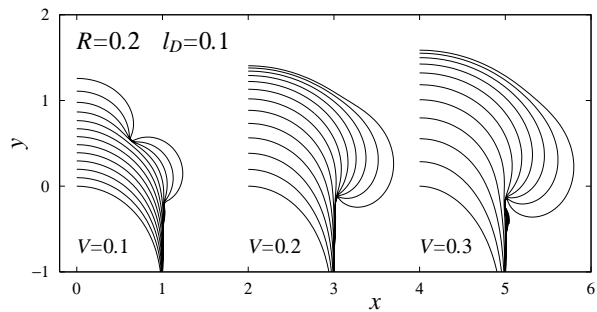}
\end{center}
\caption{
The time developments of the interface for $R=0.2$ and $\ell_D=0.1$ from
 some initial configurations.
The steady solutions of the speed $V=0.1$, 0.2, and 0.3 with
 $\ell_D=0$ are used as the initial configurations.
The time sequences are
shown with the time interval $\Delta t=1$.
}
\label{f-4}
\end{figure}

Fig.\ref{f-4} shows the results for the case of $R=0.2$ and
$\ell_D=0.1$.  The initial states are the steady solutions of $V=0.1$,
0.2, and 0.3 for $\ell_D=0$.  One may see some differences in the way
how the steady solutions are destabilized for different speeds, but
again the steady solutions are unstable for all the cases, and complex
developments of the boundary are seen after the instability.

The unstable development seen in Figs.3 and 4 shows some similarities to
the pattern found in \cite{YM00}; Concave parts of the interface evolve
eventually into cusps with the tip size of $\ell_D$, and 
the interface are pinned by them.  This leads to the
irregular development of the interface although there is no randomness
in the present model.


\section{Concluding remarks}

Before concluding, let us make some remarks.

First of all, the sweeping phenomena can be seen commonly; It is not
limited to the specific experiment we referred to, although there have
not been many controlled experiments and theoretical analyses.  Another
example may be found in a pattern formation of deposit in a drying
droplet\cite{D00}.  Actual situations varies and may not be as simple as
the one we have analyzed in this paper, but there should be a class of
phenomena characterized by the sweeping phenomenon as is discussed here.

Secondly, the boundary dynamics should be able to be
derived from the phase field model as the narrow interface width limit.
In the case of crystal growth and the viscous fingering, the interface
width is the shortest length scale in the problem, and the boundary
dynamics can be derived from the phase field model by taking appropriate
limits\cite{C89}.  On the other hand, in the sweeping
phenomenon, the length scale over which the granular density varies in
the normal direction to the interface is in the same order with the
interface width\cite{IHN05}, which complicates the formal derivation of
the boundary dynamics.

Lastly, some comments on the diffusion are in order.  We have formulated
the sweeping dynamics as the small diffusion limit, but in a real
system, some form of diffusion should exist.  We have also found that
the diffusionless model shows the short wave instability.  In
Ref.\cite{IHN05}, the ordinary diffusion is assumed within the
interface, which results in patterns with a larger scale for a slower
process; This does not seem to agree with the experiment.  On the other
hand, we regularized the model with the diffusion characterized by the
short length scale $\ell_D$, which may correspond to the grain size in
the experiment\cite{YM00}.  The diffusion along the interface is
considered to be driven by the interface motion, thus the grains do not
diffuse when the interface gets stuck, therefore, the slower interface
motion does not lead to a larger scale.


In summary, the boundary dynamics of the sweeping interface is
constructed based upon geometrical and physical analysis of the process.
To suppress the short length scale instability, the diffusion driven by
the interface motion along the interface is introduced with the short
length scale $\ell_D$. In the case of no diffusion, we obtain the
steadily propagating finger solutions for a finite range of propagation
speed, but the numerical simulations with the small diffusion suggests
that they are unstable, and a complex behavior of the interface is seen.

\begin{acknowledgments}
This work is supported by
Grant-in-Aid for Scientific Research (C) (No. 16540344) from 
JSPS, Japan.
\end{acknowledgments}


\end{document}